\documentclass[twocolumn]{aastex701}

\usepackage{hyperref,enumerate}
\usepackage{xcolor}
\usepackage{amsmath}
\usepackage{xspace}

\hypersetup{
    unicode=false,                  
    pdftoolbar=true,                
    pdfmenubar=true,                
    pdffitwindow=true,              
    pdfstartview={FitH},            
    pdftitle={HE0144-4657},         
    pdfauthor={vplacco},            
    pdfsubject={Astronomy},         
    pdfcreator={dvipdf},            
    pdfproducer={dvipdf},           
    pdfkeywords={metal-poor stars}, 
    pdfnewwindow=true,              
    colorlinks=true,                
    linkcolor=red,                  
    citecolor=blue,                 
    filecolor=magenta,              
    urlcolor=cyan,                  
    breaklinks=true,
    linktocpage
}

\newcommand{\hesstar}{HE~0144$-$4657\xspace}

\newcommand{\feh}{\ensuremath{{\rm[Fe/H]}}\xspace}

\usepackage{xcolor}

\definecolor{emerald}{rgb}{0.31, 0.78, 0.41}

\newcommand{\kmsec}{\mbox{km~s$^{\rm -1}$}}

\newcommand{\eps}[1]{\ensuremath{\log\epsilon\,(\mathrm{#1})}}
\newcommand{\vv}{{\tablenotemark{\footnotesize{a}}}}
\newcommand{\xx}{{\tablenotemark{\footnotesize{b}}}}

\newcommand{\abund}[2]{\ensuremath{[\mathrm{#1}/\mathrm{#2}]}}
\newcommand{\cfe}{\abund{C}{Fe}}
\newcommand{\nfe}{\abund{N}{Fe}}
\newcommand{\xfe}[1]{\abund{#1}{Fe}}

\newcommand{\metal}{\abund{Fe}{H}}

\newcommand{\teff}{\ensuremath{T_\mathrm{eff}}}
\newcommand{\logg}{\ensuremath{\log\,g}}
\newcommand{\Msun}{\ensuremath{M_\odot}\xspace}

\newcommand{\A}[1]{\ensuremath{A(\mathrm{#1})}}

\newcommand{\K}{\ensuremath{\mathrm{K}}}

\newcommand{\mstar}{\ensuremath{M_\star}\xspace}

\newcommand{\hesump}{\object{\mbox{HE~0144$-$4657}}}

\received{2026 January 21}
\revised{2026 March 1}
\accepted{2026 March 4}

\submitjournal{ApJL}

\begin{document}

\title{\hesstar: A Carbon-Enhanced Ultra Metal-Poor Star ($\metal \sim -4.1$) \\ from the Helmi Stream Disrupted Dwarf Galaxy\footnote{Based on observations gathered with the 6.5 m Magellan Telescopes located at Las Campanas Observatory, Chile, and the New Technology Telescope (NTT) of the European Southern Observatory (088.D-0344A), La Silla, Chile.}}


\author[0000-0003-4479-1265]{Vinicius M.\ Placco}
\affiliation{NSF NOIRLab, Tucson, AZ 85719, USA}
\email{vinicius.placco@noirlab.edu}

\author[0000-0002-9269-8287]{Guilherme Limberg}
\affiliation{Kavli Institute for Cosmological Physics, University of Chicago, 5640 S. Ellis Avenue, Chicago, IL 60637, USA}
\affiliation{Department of Astronomy \& Astrophysics, University of Chicago, 5640 S. Ellis Avenue, Chicago, IL 60637, USA}
\email{}

\author{Catherine R.\ Kennedy}
\affiliation{Randolph School, 4915 Garth Rd SE, Huntsville, AL 35802, USA}
\email{}

\author[0000-0002-4043-2727]{Norbert Christlieb}
\affiliation{Zentrum für Astronomie der Universität Heidelberg, Landessternwarte, Königstuhl 12, 69117 Heidelberg, Germany}
\email{}

\correspondingauthor{Vinicius M.\ Placco}

\begin{abstract}

We present the discovery of \hesump, an ultra metal-poor, CNO-enhanced star dynamically associated with the Helmi Stream disrupted dwarf-galaxy remnant. This star was first identified as a carbon-enhanced, metal-poor star candidate from the Hamburg/ESO objective-prism survey, then followed up with medium- and high-resolution spectroscopy. At $\metal=-4.11$, \hesump \ is the lowest metallicity star found in a stellar stream to date. Its chemistry is consistent with field halo stars in the same metallicity regime, and the light-element (atomic number $Z\leq30$) chemical abundance pattern suggests that \hesump \ is a bona-fide second-generation star with a possible Population III progenitor in the 50\,\Msun mass range with low explosion energy. One possible scenario for the origin of \hesump \ is that it was formed in an ultra-faint dwarf galaxy accreted by the Helmi Stream progenitor system before merging with the Milky Way. This discovery provides further evidence for the extragalactic origin of carbon-enhanced ultra metal-poor stars in the Milky Way and for the specific environments conducive to their formation.

\end{abstract}


\section{Introduction} 
\label{intro}

    Systematic searches for Ultra Metal-Poor \citep[UMP -- \metal\footnote{\abund{A}{B} = $\log(N_A/{}N_B)_{\star} - \log(N_A/{}N_B) _{\odot}$, where $N$ is the atom number density of an element in the star ($\star$) and the Sun ($\odot$).}$\leq-4.0$;][]{frebel2015,bonifacio2025} stars in the Milky Way have been underway for over 40 years \citep{bessell1984}. Nevertheless, only about 50 UMP stars have been identified to date. These old, low-mass stars retain in their atmospheres key information about the nature of the first (Population~III) stars formed shortly after the Big Bang \citep{klessen2023}, and are our ``local window'' into the high-redshift universe \citep{bromm2004,vanni2023}.

One of the most successful efforts to find UMP stars has been the Hamburg/ESO Survey\footnote{It is worth mentioning that HES was designed to be a wide-angle survey for bright QSOs ($12.5<B<17.5$). Coincidentally, the HES low-resolution spectroscopy turned out to be also suitable for finding low-metallicity stars in the Milky Way.} \citep[HES;][]{reimers1990,wisotzki1996}. It provided objective-prism spectra ($\mathcal{R}\sim500$) for over 4.4~million stellar sources in the southern hemisphere, with an effective area of 6726\,$\deg^2$ and a limiting magnitude of $B\sim17.5$ for stellar work \citep{christlieb2008}. High-resolution spectroscopic follow-up campaigns using HES metal-poor candidates have discovered two stars with $\metal \leq -5.0$ \citep{christlieb2002,frebel2005}, several UMP stars \citep[][]{cohen2007,norris2007,hansen2014,hansen2015,placco2016b}, and hundreds of chemically peculiar stars with $\metal >-4.0$ \citep[e.g.][among several others]{barklem2005,frebel2006,schorck2009,li2010,cohen2013,norris2013,placco2014}.

In the UMP regime, roughly 80\,\% of the stars exhibit carbon enhancement \citep{placco2014c,arentsen2022}. These Carbon-Enhanced Metal-Poor \citep[CEMP; $\cfe \geq+0.7$ --][]{aoki2007} stars at $\metal \leq-4.0$ are thought to be the direct descendants of the first stars. Theoretical models support the hypothesis that carbon is a key byproduct of the evolution of massive metal-free stars in the early universe \citep{ishigaki2014,tominaga2014}. Further observational evidence is provided by carbon measurements in Damped Ly$\alpha$ systems as early as redshift $z>2$ \citep{cooke2012,cooke2014, welsh2020}.

The next step in understanding the nature of Population~III stars is finding the astrophysical environments that would allow for their formation and evolution. Such environments would ideally retain the yields from the supernova explosions, which in turn would seed the second-generation UMP stars observed today. Connecting the chemistry of low-metallicity stars with stellar kinematics and dynamics has flourished in the era of the {\emph{Gaia}} mission \citep{gaia2016}. Through chemo-dynamical analysis, it becomes possible to associate UMP stars in the Galaxy with their potential birth environments, which include surviving Milky Way satellites such as dwarf spheroidal (dSph) and ultra-faint dwarf (UFD) galaxies, as well as disrupted dwarf galaxies in the form of stellar streams that inhabit the Galactic halo. These structures hold the key to the Milky Way's turbulent history of past interactions and mergers \citep{belokurov2013, helmi2020, deason2024gaia}.

The Helmi Stream \citep{helmi1999streams} is one such example. Believed to be the remnant of the interaction between the early Milky Way and a dwarf galaxy of similar mass to present-day surviving dSph satellites, it contains a mostly metal-poor\footnote{\citet{aguado2021} reported the discovery of a carbon-enhanced star with \metal$\sim -3$ from low-resolution spectroscopy in the Helmi Stream.} ($-2.3\leq \metal \leq-1.0$), old (11--13\,Gyr) stellar population that shares orbital properties and chemical abundances \citep{koppelman2019helmi, aguado2021, limberg2021helmistream, dodd2023subs}. Understanding its formation and origin helps us piece together the dynamical and chemical evolution of the Milky Way itself.

This article presents the discovery, spectroscopic follow-up, and chemo-dynamical analysis of \hesump, a \textit{bona fide} second-generation CNO-enhanced UMP star associated with the Helmi Stream. 
This work is outlined as follows: Section~\ref{observations} describes the target selection from HES and spectroscopic follow-up observations. The determination of stellar atmospheric parameters, chemical abundances, and kinematic/dynamical parameters is described in Section~\ref{atmparsec}. In Section~\ref{chemod}, we provide evidence of the association of \hesump\, to the Helmi Stream, as well as speculate on the possible progenitor population of this chemically-peculiar star. We also provide insights into the birth environment of second-generation stars. Our conclusions and perspectives for future work are outlined in Section~\ref{conclusion}.

\begin{deluxetable*}{llllr}
\tablecaption{Properties of \protect\hesump\label{starinfo}}
\tablewidth{0pt}
\tabletypesize{\scriptsize}
\tabletypesize{\small}
\tablehead{
\colhead{Quantity} &
\colhead{Symbol} &
\colhead{Value} &
\colhead{Units} &
\colhead{Ref.}}
\startdata
Right ascension            & $\alpha$ (J2000)    & 01:46:39.64                  & hh:mm:ss.ss       & 1  \\ 
Declination                & $\delta$ (J2000)    & $-$46:42:20.2                & dd:mm:ss.s        & 1  \\ 
Galactic longitude         & $\ell$              & 277.499                      & degrees           & 1  \\ 
Galactic latitude          & $b$                 & $-$67.602                    & degrees           & 1  \\ 
\hline                                           
Gaia DR3 ID                &                     & 4954178397218620800          &                   & 2  \\ 
Parallax                   & $\varpi$            & 0.0623 $\pm$ 0.0166          & mas               & 2  \\ 
Parallax zero point        & $\varpi_{\rm zp}$   & $-$0.0374                    & mas               & 3  \\ 
Inverse parallax distance  & $d_{\rm Gaia}$      & 10.03$^{+2.00}_{-1.43}$      & kpc               & 4    \\
Distance                   & $d_{\rm BJ}$        & 11.34$^{+0.99}_{-0.73}$      & kpc               & 5    \\
Proper motion ($\alpha$)   & PMRA                & $-$2.220 $\pm$ 0.015         & mas yr$^{-1}$     & 2  \\ 
Proper motion ($\delta$)   & PMDec               & $-$0.470 $\pm$ 0.066         & mas yr$^{-1}$     & 2  \\ 
\hline                                           
$K$ magnitude              & $K$                 & 12.691 $\pm$ 0.029           & mag               & 6  \\ 
$G$ magnitude              & $G$                 & 14.554 $\pm$ 0.003           & mag               & 2  \\ 
$BP$ magnitude             & $BP$                & 14.976 $\pm$ 0.003           & mag               & 2  \\ 
$RP$ magnitude             & $RP$                & 13.954 $\pm$ 0.004           & mag               & 2  \\ 
$V$ magnitude              & $V$                 & 14.784 $\pm$ 0.037           & mag               & 7  \\ 
$B-V$ de-reddened color    & $(B-V)_0$           & 0.67                         & mag               & 1  \\ 
$J-K$ de-reddened color    & $(J-K)_0$           & 0.51                         & mag               & 1  \\ 
Color excess               & $E(B-V)$            & 0.0123 $\pm$ 0.0001          & mag               & 8  \\ 
Bolometric correction      & BC$_V$              & $-$0.335 $\pm$ 0.049         & mag               & 9  \\ 
\hline                                           
KPHES line index           & {\texttt{KPHES}}    & 4.1                          & \AA               & 1  \\ 
GPE line index             & {\texttt{GPE}}      & 19.22                        & \AA               & 10 \\ 
EGP line index             & {\texttt{EGP}}      & $-$0.59                      & \AA               & 11 \\ 
Signal-to-noise ratio (MIKE)    @3860\AA  & S/N  & 20                           & pixel$^{-1}$      & 4  \\ 
\phantom{Signal to noise ratio (MIKE)} @4360\AA  &      & 81                    & pixel$^{-1}$      & 4  \\ 
\phantom{Signal to noise ratio (MIKE)} @5180\AA  &      & 127                   & pixel$^{-1}$      & 4  \\ 
\phantom{Signal to noise ratio (MIKE)} @6540\AA  &      & 171                   & pixel$^{-1}$      & 4  \\ 
\hline                                           
Effective Temperature      & \teff               & 5099 $\pm$ 70                & K                 & 4  \\ 
Log of surface gravity     & \logg               & 1.89 $\pm$ 0.06              & (cgs)             & 4  \\ 
Microturbulent velocity    & $\xi$               & 2.20 $\pm$ 0.20              & \kmsec            & 4  \\ 
Metallicity                & \metal              & $-$4.11 $\pm$ 0.06           & dex               & 4  \\ 
Radial velocity            & RV                  & $-$40.1 $\pm$ 1.0            & \kmsec            & 4  \\ 
\hline    
Orbital energy                  & $E$               & $(-1.34\pm0.02)\times10^5$    & ${\rm km^2\,s^{-2}}$  & 4 \\
Pericenter                      & $r_{\rm peri}$    & $8.8\pm0.3$                   & kpc   & 4 \\ 
Apocenter                       & $r_{\rm apo}$     & $16.0^{+0.8}_{-0.9}$          & kpc   & 4 \\
Eccentricity                    & $e$               & $0.29\pm0.02$                 & & 4 \\
Maximum vertical excursion      & $Z_{\rm max}$     & $12.4^{+1.0}_{-1.1}$          & kpc   & 4 \\
$x$-component of angular momentum       & $L_x$             & $1433^{+35}_{-42}$            & ${\rm kpc\,km\,s^{-1}}$   & 4 \\
$y$-component of angular momentum       & $L_y$             & $1273^{+148}_{-145}$            & ${\rm kpc\,km\,s^{-1}}$   & 4 \\
$z$-component of angular momentum       & $L_z$             & $-1570^{+34}_{-32}$            & ${\rm kpc\,km\,s^{-1}}$   & 4 \\
Perpendicular angular momentum  & $L_\perp$         & $1921^{+115}_{-128}$          & ${\rm kpc\,km\,s^{-1}}$   & 4 \\
Inclination                     & $\theta$          & $129.1^{+2.3}_{-1.8}$         & deg   & 4 \\
Radial action                   & $J_r$             & $161^{+30}_{-28}$             & ${\rm kpc\,km\,s^{-1}}$   & 4 \\
Vertical action                 & $J_z$             & $981^{+92}_{-103}$            & ${\rm kpc\,km\,s^{-1}}$   & 4 \\
%
%

\enddata
\tablerefs{
1: \citet{christlieb2008};
2:  \citet{gaia23dr3};
3:  \citet{lindegren2020};
4:  This work;
5:  \citet{Bailer-Jones+2021};
6:  \citet{skrutskie2006};
7:  \citet{henden2014};
8:  \citet{schlafly2011};
9:  \citet{casagrande2014};
10: \citet{placco2010};
11: \citet{placco2011}.
}
\end{deluxetable*}

\begin{figure*}[!ht]
 \includegraphics[width=1\linewidth]{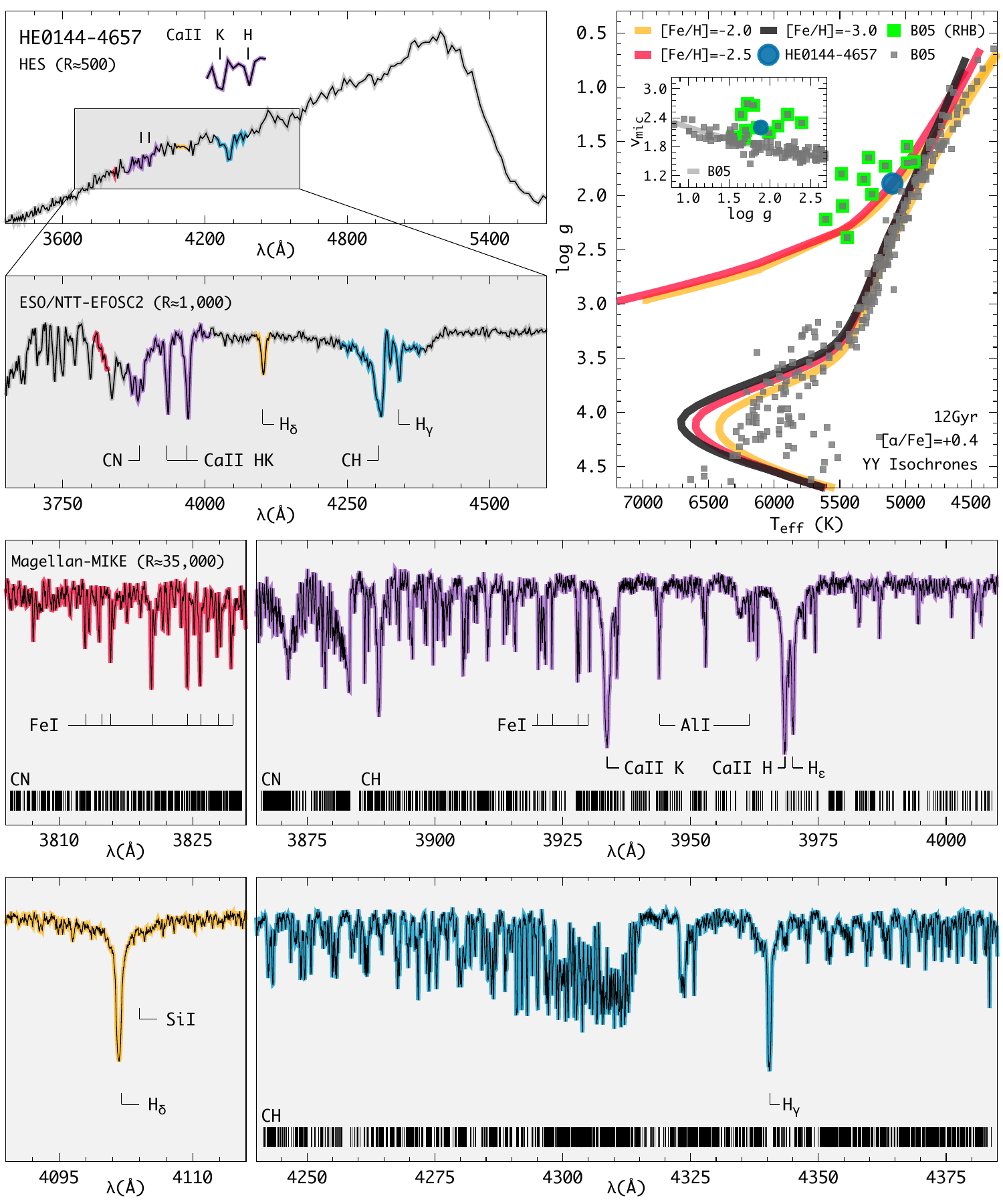}
  \caption{\protect\hesump, as observed from HES (top left), NTT/EFOSC-2 (second row from the top left), and Magellan/MIKE (bottom two rows). The different color lines reference the same regions in each spectrum, highlighting the difference in resolution. Absorption features and molecular bands of interest are labeled throughout. Top right panel: \teff\, vs. \logg\, for \hesump\, (blue circle), together with data from \citet{barklem2005} (gray squares, with green highlights for RHB stars). Solid lines show YY Isochrones from \citet{demarque2004} with 12~Gyr, 0.8\Msun, \mbox{[$\alpha$/Fe]=$+0.4$}, and \metal=$-2.0,-2.5,-3.0$. The horizontal-branch tracks were taken from \citet{dotter2008}. The inset shows $\xi$ (km/s) vs. \logg\, for the same objects.}
 \label{spectra}
\end{figure*}

\section{Target Selection and Observations}
\label{observations}

Here we briefly describe the identification, selection, and spectroscopic follow-up observations of \hesump. This star has been independently found to be metal-poor ($\metal =-2.7$) by \citet{andrae2023}, and selected as an $\metal \leq-2.0$ candidate by \citet{yao2024}, with an estimated metallicity of $\metal =-3.3$ with data-driven methods described in \citet{limberg2025}, and as a CEMP candidate by \citet{lucey2023}. Table~\ref{starinfo} lists basic information and derived quantities for \hesump, measured in this work and other studies in the literature.

\subsection{Hamburg/ESO Survey}

\hesump\, was first identified as a metal-poor star candidate by \citet{christlieb2008} in the HES digitized objective-prism plates. Candidates were selected by their position in the $(B-V)_0$ vs. {\texttt{KPHES}}\footnote{The {\texttt{KPHES}} line index measures the strength of the \ion{Ca}{2}~K line at 3933.66\AA. See Table 1 in \citet{christlieb2008}.} and $(J-K)_0$ vs. {\texttt{KPHES}} diagrams, then the 1D spectra were subject to visual inspection to reject false positives and apply a subjective ranking based on the presence and strength of the Ca{\sc{ii}}~K line. \hesump\, was classified as {\texttt{mpcc}}\footnote{``A significant Ca{\sc{ii}}~K line appears against the continuum.''} with an estimated $\metal=-2.8$ and $\cfe=1.0$. The star was then added to spectroscopic follow-up target lists for further characterization.

In addition, \hesump\, was independently identified by \citet{placco2010,placco2011}, who selected metal-poor candidates assuming that an increasing fraction (with decreasing metallicity) of those would be carbon-enhanced \citep{rossi1999, lucatello2006}. This was done via the {\texttt{GPE}} and {\texttt{EGP}} line indices\footnote{The {\texttt{GPE}} \citep[\texttt{GPHES} Extended;][]{placco2010} and {\texttt{EGP}} \citep[Extended \texttt{GPHES};][]{placco2011} line indices are based on the \texttt{GPHES} line index from \citet{christlieb2008}, which measures the strength of the CH G-band centered around 4300\AA.}, which flagged \hesump\, as potentially interesting due the apparent weakness of the \ion{Ca}{2} lines and strength of the carbon G-band.
The upper left panel of Figure~\ref{spectra} shows the HES spectrum of \hesump, highlighting selected regions of interest. Of note are the \ion{Ca}{2}~H and K absorption features (purple line and inset) and the carbon G-band (light blue line). Even at the low resolution of the HES spectra ($\mathcal{R} \sim500$), it is already possible to infer that \hesump\, is metal-poor and carbon-enhanced. 

\subsection{Medium-resolution Spectroscopy}

 The medium-resolution spectrum of \hesump\, was obtained on 2011 October 16 with the EFOSC-2 spectrograph \citep{buzzoni1984} mounted on the 3.5~m ESO New Technology Telescope (NTT). Grism\#7 (600 gr~mm$^{-1}$) and a 1\farcs0 slit with 1$\times$1 binning were used, resulting in a wavelength coverage of [3450:5100]~\AA\, and resolving power of $\mathcal{R}\sim1000$. With a 600-second exposure time, a signal-to-noise ratio $S/N \sim50$ at 4000\,\AA \ was achieved. The data reduction was performed using standard {\texttt{IRAF}}\footnote{NOIRLab {\tt IRAF} is distributed by the Community Science and Data Center at NSF NOIRLab, which is managed by the Association of Universities for Research in Astronomy (AURA) under a cooperative agreement with the U.S. National Science Foundation.} routines for long-slit spectroscopy.

A section of the NTT/EFOSC-2 spectrum is shown in the second panel from the top left of Figure~\ref{spectra}, highlighting regions and absorption features of interest. The narrow \ion{Ca}{2} lines, coupled with strong CN and CH molecular features, suggested a CEMP status for \hesump. A preliminary analysis using the methods outlined in \citet{lee2013} and \citet{placco2018} provided stellar atmospheric parameters ($\teff = 5283\,{\rm K}$, $\logg=2.05$, and $\metal=-3.83$) and carbon abundance ($\cfe=2.76$) consistent with a UMP classification, which motivated the high-resolution spectroscopic follow-up.

\subsection{High-resolution Spectroscopy}

The high-resolution spectra for \hesump\, were obtained on 2012 October 3 and 2025 October 15 with the Magellan Inamori Kyocera Echelle \citep[MIKE;][]{mike} spectrograph mounted on the 6.5~m Magellan-Clay Telescope at Las Campanas Observatory. The observing setup included a 0\farcs7 slit with $2\times2$ binning, yielding a resolving power of $\mathcal{R}\sim 28,000$ in the blue spectrum and $\mathcal{R}\sim 22,000$ in the red spectrum. The wavelength coverage of the combined blue and red spectra is [3350:8500]\,\AA. The data was reduced using the {\texttt{CarPy}}\footnote{\href{https://code.obs.carnegiescience.edu/pipelines/mike}{https://code.obs.carnegiescience.edu/pipelines/mike}} software \citep{kelson2003}. Table~\ref{starinfo} lists the $S/N$ measured at various regions of the reduced spectra.

The bottom two rows in Figure~\ref{spectra} show selected regions of the MIKE spectrum, highlighting features of interest for the determination of stellar parameters and chemical abundances (see Section~\ref{atmparsec}). As expected, \hesump\, shows very strong CN and CH molecular features from 3800\,\AA \ to 4400\,\AA, which poses a challenge for measuring absorption features of other species due to line blending. At the same time, these molecular features allow for the determination of accurate carbon and nitrogen abundances, as described in the following Section.

\section{Methods}
\label{atmparsec}

\subsection{Atmospheric Parameters}

The stellar atmospheric parameters (\teff, \logg, \metal, and $\xi$) for \hesump\, were determined using photometric inputs, listed in Table~\ref{starinfo}, and the MIKE spectrum.
The effective temperature (\teff) was calculated from the color-\teff-\metal\, relations for red giant-branch stars derived by \citet{mucciarelli21}, using the $G$, $BP$, and $RP$ magnitudes from Gaia DR3 \citep{gaia23dr3}, and the $K$ magnitudes from 2MASS \citep{skrutskie2006}, with reddening corrections from \citet{gaia2018} and \citet{mccall2004}. 
From the magnitudes and their uncertainties, 10$^5$ samples were drawn assuming normal distributions. For each input color ($BP-RP$, $BP-G$, $G-RP$, $BP-K$, $RP-K$, and $G-K$), the median temperatures were calculated assuming $\metal=-4.0$. The final value ($\teff=5099\pm70$) is the weighted mean of the median temperatures for the six different colors.
The surface gravity (\logg) was determined from fundamental relations \citep[e.g., Equation 1 in][]{roederer2018} by drawing $10^5$ samples from the input parameters in Table~\ref{starinfo}. The median of those calculations is the adopted value ($\logg=1.89\pm0.06$), with the uncertainty given by their standard deviation.

The iron-abundance metallicity for \hesump\, ($\metal=-4.11\pm0.06$) was determined spectroscopically from the equivalent widths (EWs) of \ion{Fe}{1} absorption lines in the MIKE spectra, by fixing the \teff\, and \logg. The EWs were obtained by fitting Gaussian profiles to the observed features using the {\texttt{Spectroscopy Made Harder}} \citep[\texttt{SMHr};][]{casey2014} software. Due to the presence of strong CN and CH molecular features between 3800\,\AA \ and 4400\,\AA \ (see bottom panels of Figure~\ref{spectra}), only 25 reliable \ion{Fe}{1} lines could be measured, along with two \ion{Fe}{2} lines. 
\texttt{SMHr} calculates \metal\, using the 2017 version of the \texttt{MOOG}\footnote{\href{https://github.com/alexji/moog17scat}{https://github.com/alexji/moog17scat}} code \citep{sneden1973,sobeck2011}, employing one-dimensional plane-parallel model atmospheres with no overshooting \citep{castelli2004}, assuming local thermodynamic equilibrium (LTE).
The microturbulent velocity ($\xi=2.20\pm0.20$) was determined by minimizing the trend between the reduced equivalent width $\log(\rm EW/\lambda)$ and abundances for \ion{Fe}{1} lines. The final stellar parameters for \hesump\, are listed in Table~\ref{starinfo}.

The top right panel of Figure~\ref{spectra} shows the position of \hesump\, (blue circle) in the Kiel diagram, together with the sample from \citet{barklem2005} as gray squares (red horizontal branch -- RHB -- stars are highlighted with a green background). The Yonsei-Yale (YY) Isochrones from \citet{demarque2004} with 12\,Gyr, 0.8\,\Msun, \mbox{$[\alpha/{\rm Fe}]=+0.4$}, and $\metal=-2.0$, $-$2.5, and $-$3.0 are shown as solid lines. The horizontal-branch tracks were taken from \citet{dotter2008}. From its position with respect to the isochrones, \hesump\, can be classified as an RHB star. The inset in the top right panel of Figure~\ref{spectra} shows the $\xi$ vs. \logg\, distribution of the stars from \citet{barklem2005} and \hesump, which shows consistent values for the RHB stars. Also shown is the B05 quadratic relation from \citet{ji2023}.

\begin{figure*}
 \includegraphics[width=1\linewidth]{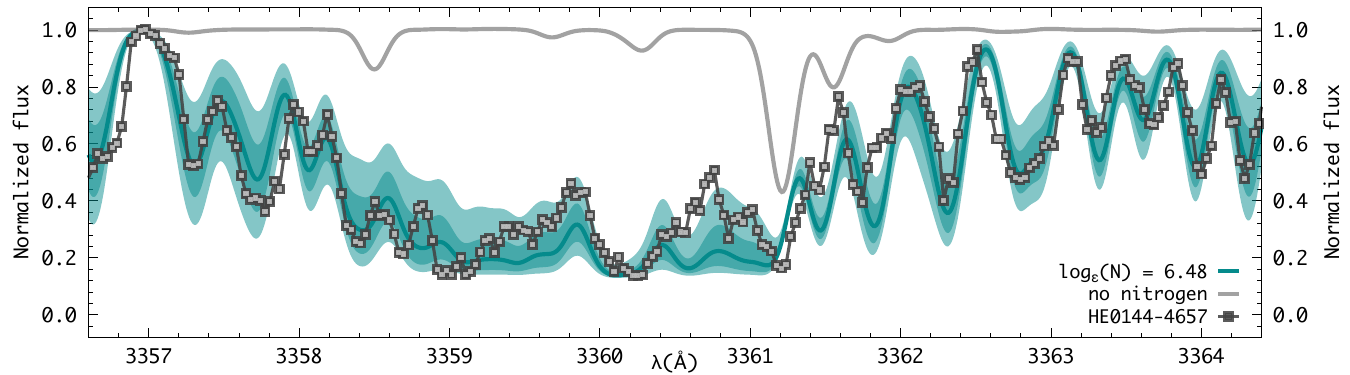}
 \includegraphics[width=1\linewidth]{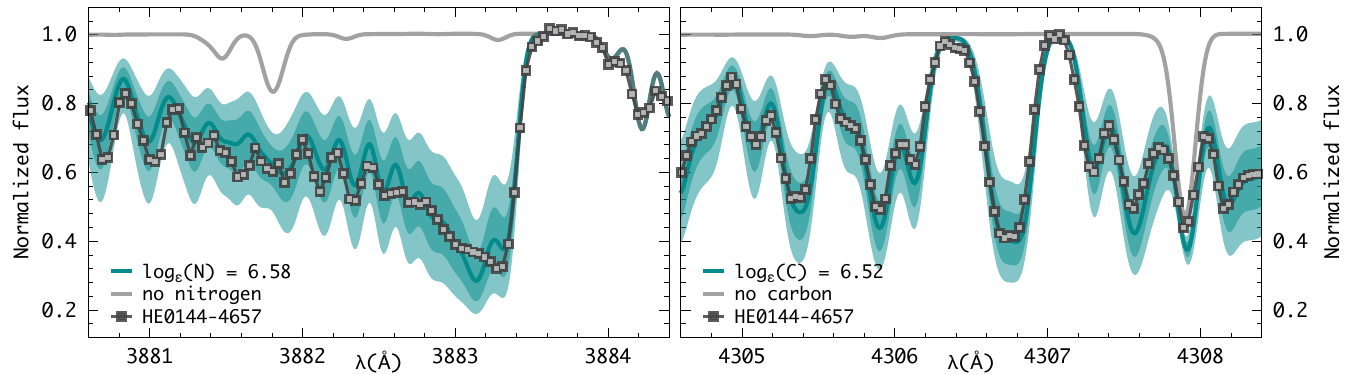}
 \includegraphics[width=1\linewidth]{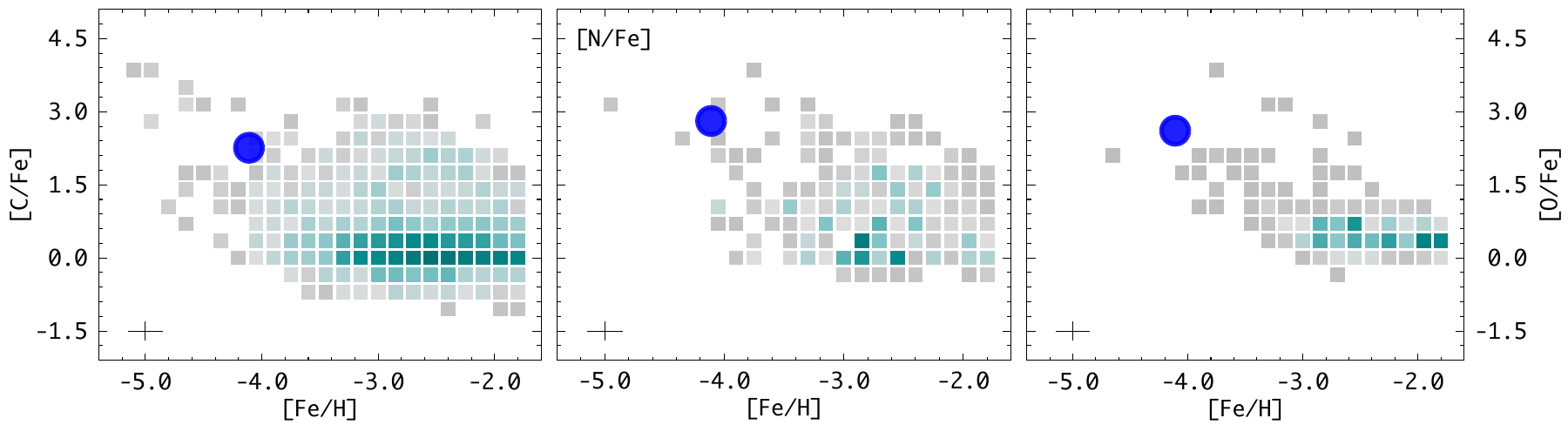}
 \includegraphics[width=1\linewidth]{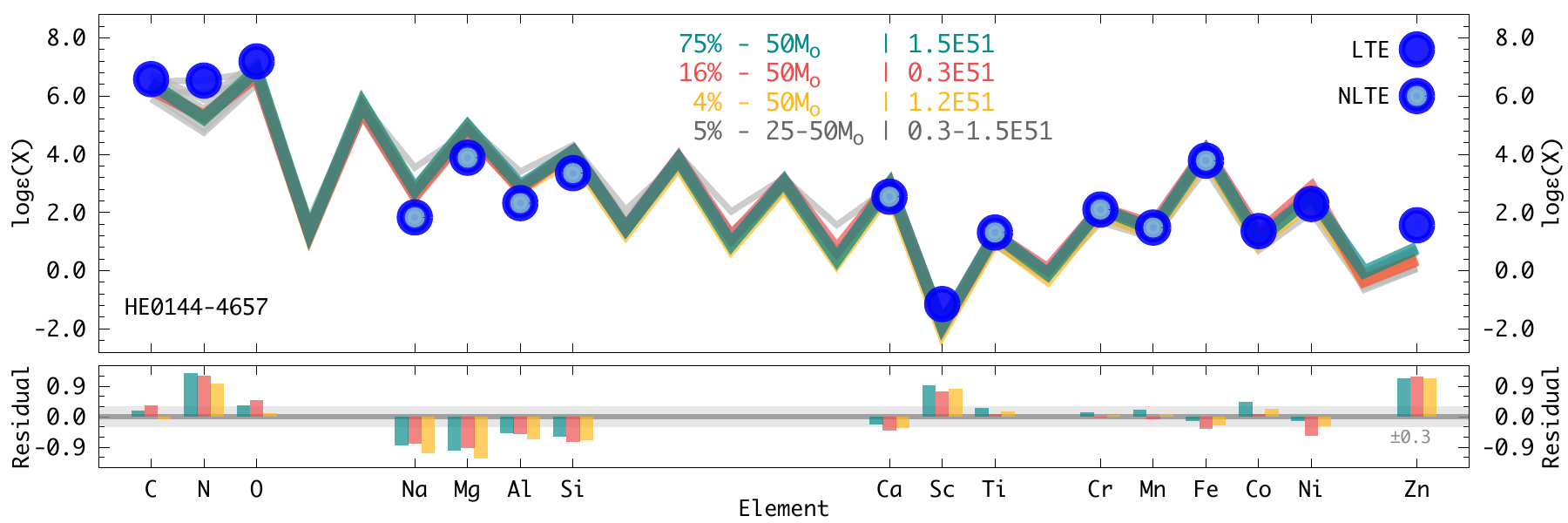}
     \caption{First and second rows: Abundance determination via spectral synthesis for nitrogen and carbon. The filled gray squares connected by the black line represent the MIKE spectrum, the teal line is the best fit, and the shaded regions represent $\pm0.15$ and $\pm0.30$~dex from the best-fit abundance. Also shown is a synthetic spectrum without carbon and nitrogen (gray line). Third row from the top: Selected LTE [X/Fe] abundance ratios as a function of \metal \ for \hesump \ (blue circle), compared with the \texttt{JINAbase} \citep{jinabase} and \texttt{SAGA} \citep{saga2008} literature compilations (density map). Additional references include \citet{limberg2025,placco2025}. Bottom panel: Light-element abundance pattern for \hesump \ (LTE and NLTE - blue points), compared with Population~III supernova yields (solid lines, residuals given as vertical colored bars). See text for details.}
 \label{csyn}
\end{figure*}

\subsection{Chemical Abundances}

From the MIKE spectrum, chemical abundances for 18 elements were determined for \hesump, using both EWs and spectral synthesis\footnote{The best fit abundances were determined using the minimization technique described in \citet{placco2024}.}. The line lists were generated by the \texttt{linemake} code\footnote{\href{https://github.com/vmplacco/linemake}{https://github.com/vmplacco/linemake/}} \citep{placco2021}. Logarithmic number abundances ($\log\epsilon$(X)) and abundance ratios (\xfe{X}) adopt the solar photospheric abundances ($\log\epsilon_{\odot}$\,(X)) from \citet{asplund2009}. Final abundance values and the number of lines measured ($N$) for each element are given in Table~\ref{abund}. The $\sigma$ values represent the adopted uncertainty\footnote{For abundances with standard deviations $<0.10$ and $N>1$ in Table~\ref{abund}, we set a standard minimum value of $\sigma=0.10$. When $N=1$ for the EW analysis, $\sigma_a=0.15$. For the abundances determined through spectral synthesis, uncertainties were estimated by minimizing the residuals between the MIKE data and a set of synthetic spectra.}.

For the carbon abundance determination, the CH G-band was synthesized in two regions: 4300--4315\,\AA \ and 4320--4326\,\AA, both with a best-fit abundance of $\eps{C}=6.52$. The $^{12}{\rm C}/^{13}{\rm C}=4$ isotopic ratio was determined by fixing the carbon abundance found for the G-band and synthesizing different isotopic ratios in the 4217--4220\,\AA \ region. The nitrogen abundance was determined from two molecular features: NH at 3360\,\AA \ ($\eps{N}=6.48$) and CN at 3883\,\AA \ ($\eps{N}=6.58$, using a fixed $\eps{C}=6.52$). The first two rows of Figure~\ref{csyn} show the spectral syntheses of the NH, CN, and CH molecular features. The filled gray squares connected by the black line represent the MIKE spectrum, the teal line is the best fit, and the shaded regions represent $\pm0.15$ and $\pm0.30$~dex from the best-fit abundance, which is used to estimate the uncertainties. The gray lines show synthetic spectra without carbon and nitrogen. In all cases, there is an excellent agreement between observations and synthetic data.

For the remaining elements listed in Table~\ref{abund}, abundances for \ion{O}{1} (7774\,\AA), \ion{Al}{1} (3944\,\AA \ and 3961\,\AA), \ion{Si}{1} (3905\,\AA), \ion{Ca}{1} (4226\,\AA), \ion{Sc}{2} (4246\,\AA), \ion{Cr}{1} (4274\,\AA \ and 4289\,\AA), \ion{Mn}{1} (4030\,\AA \ and 4033\,\AA), \ion{Zn}{1} (4722\,\AA), \ion{Sr}{2} (4077\,\AA \ and 4215\,\AA), and \ion{Ba}{2} (4554\,\AA) were determined through spectral synthesis and \ion{Na}{1}, \ion{Mg}{1}, \ion{Ti}{2}, \ion{Co}{1}, and \ion{Ni}{1} through EW analysis.
Non-LTE (NLTE) corrections were obtained for selected elements using the INSPECT\footnote{\href{http://www.inspect-stars.com/}{http://www.inspect-stars.com/}} database (\ion{Na}{1}), \citet{nordlander2017b} (\ion{Al}{1}), and the MPIA NLTE\footnote{\href{https://nlte.mpia.de/}{https://nlte.mpia.de/}} \citep{mpia} database (\ion{Mg}{1}, \ion{Si}{1}, \ion{Ca}{1}, \ion{Ti}{2}, \ion{Cr}{1}, \ion{Mn}{1}, \ion{Fe}{1}, and \ion{Fe}{2}). The average NLTE corrections are listed in the $\Delta{\rm NLTE}$ column in Table~\ref{abund}. Some elements are heavily affected by NLTE effects, with the largest corrections for \ion{Cr}{1} ($+0.92$\,dex), \ion{Mn}{1} ($+0.85$\,dex), and \ion{Al}{1} ($+0.80$\,dex). 

\begin{deluxetable}{lrrrccr}[!ht] 
\tabletypesize{\footnotesize}
\tablewidth{0pc}
\tablecaption{Abundances for Individual Species \label{abund}}
\tablehead{
\colhead{Ion}                         & 
\colhead{$\log\epsilon_{\odot}$\,(X)} & 
\colhead{$\log\epsilon$\,(X)}         & 
\colhead{$\mbox{[X/Fe]}$}             & 
\colhead{$\Delta$NLTE\xx}             & 
\colhead{$\sigma$}                    & 
\colhead{$N$}                         }
\startdata
C           &    8.43 &    6.52 &    2.20 & \nodata &    0.20 &       2 \\ 
C\vv        &    8.43 &    6.58 &    2.26 & \nodata &    0.20 &       2 \\ 
N           &    7.83 &    6.53 &    2.81 & \nodata &    0.20 &       2 \\ 
\ion{O}{1}  &    8.69 &    7.19 &    2.61 & \nodata &    0.25 &       1 \\ 
\ion{Na}{1} &    6.24 &    1.79 & $-$0.34 &    0.04 &    0.10 &       2 \\ 
\ion{Mg}{1} &    7.60 &    3.62 &    0.13 &    0.26 &    0.10 &       2 \\ 
\ion{Al}{1} &    6.45 &    1.52 & $-$0.82 &    0.80 &    0.15 &       2 \\ 
\ion{Si}{1} &    7.51 &    3.26 & $-$0.14 &    0.09 &    0.15 &       1 \\ 
\ion{Ca}{1} &    6.34 &    2.19 & $-$0.04 &    0.35 &    0.15 &       1 \\ 
\ion{Sc}{2} &    3.15 & $-$1.15 & $-$0.19 & \nodata &    0.15 &       1 \\ 
\ion{Ti}{2} &    4.95 &    1.24 &    0.40 &    0.07 &    0.10 &       8 \\ 
\ion{Cr}{1} &    5.64 &    1.18 & $-$0.35 &    0.92 &    0.15 &       2 \\ 
\ion{Mn}{1} &    5.43 &    0.63 & $-$0.69 &    0.85 &    0.10 &       2 \\ 
\ion{Fe}{1} &    7.50 &    3.39 &    0.00 &    0.39 &    0.10 &      25 \\ 
\ion{Fe}{2} &    7.50 &    3.40 &    0.01 &    0.02 &    0.10 &       2 \\ 
\ion{Co}{1} &    4.99 &    1.36 &    0.48 & \nodata &    0.15 &       1 \\ 
\ion{Ni}{1} &    6.22 &    2.29 &    0.18 & \nodata &    0.10 &       2 \\ 
\ion{Zn}{1} &    4.56 &    1.56 &    1.11 & \nodata &    0.20 &       1 \\ 
\ion{Sr}{2} &    2.87 & $-$1.54 & $-$0.30 & \nodata &    0.20 &       2 \\ 
\ion{Ba}{2} &    2.18 & $-$2.82 & $-$0.89 & \nodata &    0.20 &       1 \\ 
\enddata
\tablenotetext{a}{Evolutionary corrections from \citet{placco2014c}.}
\tablenotetext{b}{Assuming \metal=$-4.0$ for \ion{Na}{1}.}
\tablerefs{
NLTE corrections -- 
\ion{Na}{1}: \citet{lind2011};
\ion{Mg}{1}: \citet{bergemann2015};
\ion{Al}{1}: \citet{nordlander2017b};
\ion{Si}{1}: \citet{bergemann2013};
\ion{Ca}{1}: \citet{mashonkina2007};
\ion{Ti}{2}: \citet{bergemann2011};
\ion{Cr}{1}: \citet{bergemann2010};
\ion{Mn}{1}: \citet{bergemann2019};
\ion{Fe}{1}: \citet{bergemann2012b}.
}
\end{deluxetable}

\subsection{Kinematics and Dynamics}
\label{kindyn}

With the available phase-space information (Table \ref{starinfo}), we compute kinematics in the Cartesian Galactocentric frame for \hesstar. We adopt the position and velocity of the Sun relative to the Galactic center to be $(X,Y,Z)_\odot=(-8.2, 0.0, 0.0)$\,kpc \citep{hawthorn2016} and $(V_x,V_y,V_z)_\odot=(11.10, 245.04, 7.25)\,\kmsec$ \citep{Schonrich+2010, mcmillan2017}, respectively. In this reference frame, $L_z < 0$ signifies prograde motion, where $L_z$ is the (vertical) $z$-component of angular momentum. 

We integrate \hesstar's complete orbit for 20\,Gyr forward in the axisymmetric Milky Way model potential of \citet{mcmillan2017} using the \texttt{AGAMA} package \citep{agama}. We perform 1,000 realizations of this star's orbit from the Gaussian uncertainties in its phase-space quantities. The nominal values for the kinematic/dynamical properties are taken as the medians of the resulting distributions, and the 16$^{\rm th}$/84$^{\rm th}$ percentiles are used as lower/upper bound uncertainties (see Table \ref{starinfo} for selected parameters of interest). For comparison with \hesstar, we apply this same procedure to a stellar sample from the Sloan Extension for Galactic Understanding and Exploration (SEGUE) survey \citep{yanny2009} with available spectro-photo-astrometric distances from the \texttt{StarHorse} code \citep{queiroz2023sh}, which is particularly suited for the study of the distant metal-poor Milky Way halo \citep{limberg2023}.


\begin{figure*}[pt!]
\centering
\includegraphics[width=1.4\columnwidth]{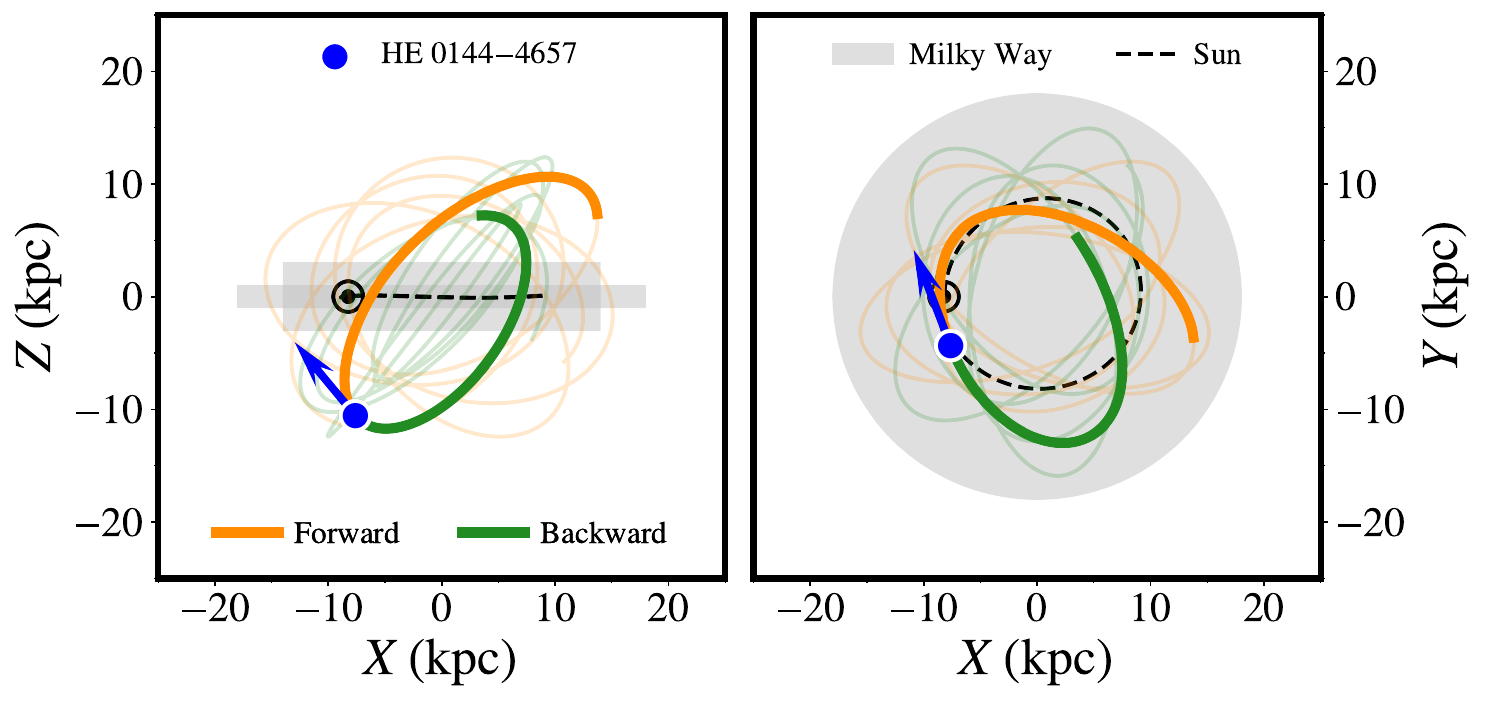}
\includegraphics[width=2.1\columnwidth]{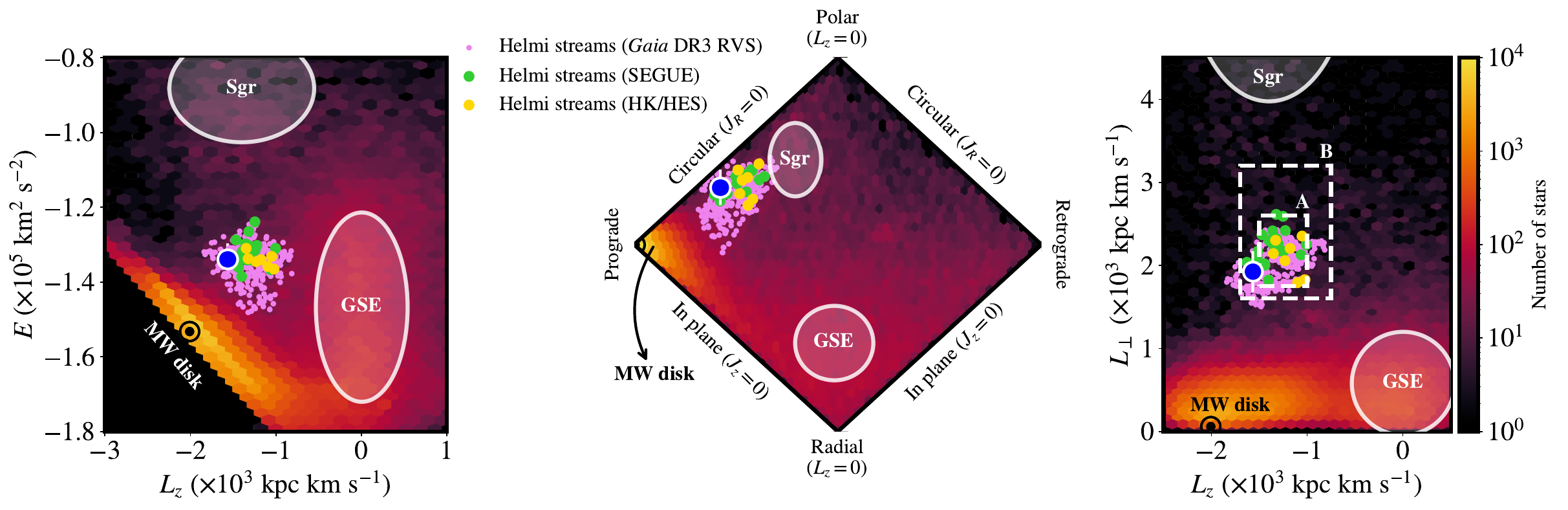}
\caption{Top panels: fiducial orbit for \hesstar (blue circle) in Cartesian Galactocentric coordinates within the \citet{mcmillan2017} model potential. Top left: $(X,Z)$; MW edge-on view. Top right: $(X,Y)$; face-on. Blue arrows exhibit the present-day velocity vector of \hesstar. Orange and green lines are forward and backward trajectories, respectively. Thicker lines show 200\,Myr integration, while thinner transparent lines represent 2\,Gyr total. Black dashed lines are the orbit of the Sun with the same assumptions (Section \ref{kindyn}) and the black symbol marks its current Galactic location. The gray regions illustrate the MW's thin disk with 18\,kpc radius and 1\,kpc maximum height. The thicker portion in the top left panel is the thick disk (14\,kpc radius and 3\,kpc height). Bottom panels: dynamical parameter spaces. Bottom left: $(L_z,E)$. Bottom center: projected action space (or action `diamond'). The horizontal axis displays $L_z/J_{\rm total} \in [-1,+1]$, where $J_{\rm total} = J_R + |L_z| + J_z$ (see text). Vertical axis show $(J_z - J_R)/J_{\rm total} \in [-1,+1]$. Bottom right: $(L_z,L_\perp)$, where $L_\perp = \sqrt{L_x^2 + L_y^2}$. The white dashed rectangles delineate the Helmi Stream kinematic locus. In all bottom panels, green, yellow, and pink dots are Helmi Stream members from the SEGUE survey \citep{myeong2018shards}, a very metal-poor sample ($\feh\leq-1.8$) including HES data \citep[][]{limberg2021b}, and Gaia DR3 RVS instrument \citep{dodd2023subs}. We also highlight the dynamical volume occupied by the most relevant halo substructures, namely Gaia-Sausage/Enceladus \citep[``GSE'';][]{belokurov2018, helmi2018} and Sagittarius stream \citep[``Sgr'';][]{ibata1994, majewski2003sgr}. The colored background 2D histograms consist of the SEGUE/\texttt{StarHorse} catalog \citep{limberg2023, queiroz2023sh}.
\label{orbs}}
\end{figure*}

\section{Analysis and Discussion}
\label{chemod}

\subsection{Dynamical Association with the Helmi Stream}
\label{hstr}

Kinematics in Table \ref{starinfo} and the top panels in Figure \ref{orbs} reveal that \hesstar is clearly on a halo-like orbit. \hesstar's present-day location is at an $(X,Y)$-projected distance from the Sun of just $\sim$4\,kpc, but its vertical displacement is $>$10\,kpc from the Galactic plane. Indeed, the maximum vertical excursion achieved by \hesstar is $Z_{\rm max} > 12\,{\rm kpc}$ in the \citet{mcmillan2017} model potential. However, unlike stars from the dwarf-galaxy merger remnant Gaia-Sausage/Enceladus \citep[$L_z \sim 0$ in bottom left panel of Figure \ref{orbs};][]{belokurov2018, haywood2018, helmi2018}, \hesstar is on a polar orbit with low eccentricity ($e \sim 0.3$\footnote{$e = (r_{\rm apo}-r_{\rm peri})/(r_{\rm apo}+r_{\rm peri})$, where $r_{\rm apo}$ and $r_{\rm peri}$ are the orbital apocenter and pericenter, respectively.}). We further discard an association with the Sagittarius stream \citep{ibata1994, majewski2003sgr} since this substructure's stellar population displays a mean orbital energy much higher than \hesstar. Also, the $y$-component of angular momentum of \hesstar ($L_y \sim 1270\,{\rm kpc\,km\,s^{-1}}$) has the opposite sign of Sagittarius stream stars' \citep[$L_{y,{\rm Sgr}} \lesssim -3000\,{\rm kpc\,km\,s^{-1}}$;][]{limberg2023}. Lastly, we reject membership within the Wukong/LMS-1 stream \citep{naidu2020subs, yuan2020lms1} since \hesstar has a more extreme $L_z \sim -1570\,{\rm kpc\,km\,s^{-1}}$ \citep[compared to $L_{z,{\rm Wuk}} > -1000\,{\rm kpc\,km\,s^{-1}}$;][]{naidu2020subs, limberg2024wuk} and milder orbital inclination $\theta \sim 130^\circ$\footnote{$\theta = \arccos{(L_z/L)}$, where $L = \sqrt{L_x^2 + L_y^2 + L_z^2}$. Note that, for prograde circular orbits $L \sim -L_z$, hence $\theta \sim 180^\circ$ for stars in the Milky Way disk such as the Sun ($\theta_\odot \sim 178^\circ$).} \citep[versus $\theta_{\rm Wuk} < 120^\circ$;][]{johnson2023wukgse, limberg2024wuk}.

In the spirit of the above discussion, we find the kinematics of \hesstar to be fully consistent with the Helmi Stream \citep{helmi1999streams}, a halo substructure of dwarf-galaxy origin with progenitor stellar mass of $M_\star \sim 5 \times 10^7\,M_\odot$\footnote{Assuming a mean metallicity of $\langle\rm[Fe/H]\rangle \sim -1.5$ \citep{dodd2023subs} and accretion redshift between 0.5 and 1.0 \citep{koppelman2019helmi} applied to the Local Group stellar mass-metallicity relation \citep{kirby2013mzr} with a redshift evolution prescription \citep{ma2016mzr}.} (similar to surviving Milky Way satellites Fornax dSph and Sagittarius dSph). \hesstar fulfills the angular-momentum criteria from \citet[][bottom right panel in Figure \ref{orbs}]{koppelman2019helmi} which includes $L_z$ and $L_\perp = \sqrt{L_x^2 + L_y^2}$ components, as well as the orbital eccentricity and inclination recommended values by \citet[][see also \citealt{aguado2021}]{limberg2021helmistream}. For comparison with \hesstar, we compiled Helmi Stream samples from the literature with member stars obtained exclusively from data-driven algorithms \citep{myeong2018shards, limberg2021b, dodd2023subs}\footnote{Since the \citet{myeong2018shards} Helmi Stream members come from SEGUE, we simply adopt \texttt{StarHorse} distances and pre-computed dynamical parameters (Section \ref{kindyn}). For the \citet{limberg2021b} and \citet{dodd2023subs} samples, we adopt \citet{Bailer-Jones+2021} distances and integrate orbits in the \citet{mcmillan2017} potential as in Section \ref{kindyn}.}. Left and middle panels in the bottom row of Figure \ref{orbs} show the compatibility between \hesstar and these independent Helmi Stream samples in $(L_z, E)$ and projected action space, where $\mathbf{J}=(J_R, L_z, J_z)$ is the action vector in cylindrical frame, respectively, corroborating this association.

\subsection{Insights on Progenitor Population and Comparison with Literature Abundances}
\label{literature}

We compared the abundances for carbon, nitrogen, and oxygen measured for \hesump\, with data from the \texttt{JINAbase}\footnote{\href{https://github.com/Mohammad-Mardini/JINAbase-updated}{https://github.com/Mohammad-Mardini/JINAbase-updated}} \citep{jinabase} and \texttt{SAGA}\footnote{\href{http://sagadatabase.jp/}{http://sagadatabase.jp/}} \citep{saga2008} literature compilations. The comparison is restricted to the metallicity range $-5.2\leq  \metal \leq-1.8$, and the results are shown in the panels on the third row of Figure~\ref{csyn}. The CNO abundances for \hesstar are in agreement with the general trends in the literature data for the UMP regime. With $\abund{C}{N}=-0.61$, we infer that \hesstar has experienced mixing \citep{spite2005,schichtel2025}, consistent with its extremely high nitrogen abundance (\nfe=$+2.81$), RHB evolutionary status, and location in the Kiel diagram (top-right panel in Figure~\ref{spectra}).

Due to its low metallicity and high carbon abundance, \hesstar is a prime candidate to be a true second-generation star, formed from interstellar gas enriched by a single supernova explosion. 
With \eps{C}=A(C)=6.58 and subsolar \xfe{Ba}, \hesstar can be placed in the low carbon band defined by \citet{spite2013} and classified as a Group~III CEMP-no \citep{yoon2016}. The radial velocity variation between the two high-resolution spectra is within uncertainties, suggesting \hesstar is not in a binary system, which corroborates with the $17\pm9\%$ binary fraction among CEMP-no stars determined by \citet{hansen2016}. We note, however, the case of HE~0107$-$5240, a \metal$<-5$ CEMP-no in a long-period binary system \citep{caffau2025}. Further radial velocity monitoring for \hesstar would be desirable to further assess its binary status.

\citet{hartwig2018} proposed \abund{Mg}{C} as a chemical marker to distinguish mono- vs. multi-enriched stars. With $\abund{Mg}{C}=-2.07$, \hesstar is well within the mono-enriched regime. More recently, \citet{hartwig2023} developed a data-driven method to classify UMP stars into mono- or multi-enriched stars with support vector machines. We used the {\texttt{emu-c}} code\footnote{\href{https://gitlab.com/thartwig/emu-c}{https://gitlab.com/thartwig/emu-c}}, which takes abundances for C, O, Na, Mg, Al, Si, Ca, Cr, Mn, Fe, Co, Ni, and Zn as inputs. From the LTE abundances in Table~\ref{abund}, the probability of \hesump\, being mono-enriched is $86\pm8$\%. Using the NLTE abundances when available, the probability increases to $89\pm7$\%.

To quantify the mass and explosion energy of the potential progenitor of \hesump, we used the \texttt{starfit}\footnote{\href{https://starfit.org/}{https://starfit.org/}} tool, which contains nucleosynthesis yields\footnote{In this work, we have used the \texttt{znuc2012.S4} model database.} for 16,800 Population~III faint-supernova \citep{heger2002,heger2010} models with masses in the 9.6~\Msun to 100~\Msun range, explosion energies 0.3--$10\times10^{51}$~erg, and a mixing parameter (0.0 to 0.25). For the fitting exercise, we resampled the light-element ($Z\leq30$) abundance pattern for \hesump\, 10,000 times, assuming normal distributions for each element, centered on the observed abundance and with dispersion $\sigma$. The results are shown in the bottom panel of Figure~\ref{csyn}. The observed abundances (using the NLTE corrections when available) are shown as blue circles, and the best-fit models as solid lines, colored by percentage occurrence, mass, and explosion energy. 99.2\% of the re-sampled abundance patterns have their best-fit model with 50\,\Msun and $E_{\rm SN}\leq1.8\times10^{51}$\,erg.

To test the robustness of this solution, we repeated the exercise above with the following changes: (i) using LTE abundances only; (ii) removing Zn; (iii) removing Zn and O; (iv) removing N; (v) using \eps{N}$-0.5$; (vi) using \eps{N}$-1.0$; and (vii) using \eps{C}$+0.30$ and \eps{N}$-1.50$\footnote{These changes aim to emulate the natal carbon and nitrogen abundances of \hesump, assuming a 0.30\,dex carbon depletion at the tip of the red giant branch, which translates into an 1.50\,dex increase in the nitrogen abundance to maintain the \xfe{(C+N)} abundance ratio constant.}. The only case where the most frequent best-fit model was not a low-energy 50\,\Msun\, is for (iv), removing nitrogen, with progenitor masses of $\sim$11\,$\Msun$. The importance of nitrogen in constraining the progenitor mass for \hesstar agrees with the conclusions found by \citet{frebel2015b,placco2015,placco2016b}, and supports the hypothesis that \hesump\, is a genuine second-generation star formed from a gas cloud enriched by a single low-energy 50\,\Msun Population~III star in the early universe.

\subsection{Implications on the Birth Environment of Second-generation CEMP Stars}
\label{cemp}

There is accumulating evidence that the CEMP stars fraction in relatively bright dwarf galaxies is lower than in both the Milky Way and UFD systems \citep[e.g.,][]{lucchesi2024}. This observational fact extends from Sculptor dSph \citep[$\mstar \sim 10^6\,\Msun$;][]{chiti2018scl, skuladottir2024} to the most luminous Milky Way satellite, the Large Magellanic Cloud \citep[LMC, $\mstar \sim 10^9\,\Msun$;][]{chiti2024lmc}. In fact, while the CEMP fraction is $\sim$80\% in the Milky Way at $\metal < -4.0$ \citep{placco2014c,arentsen2022} and reaches 100\% in UFDs already at $\metal \sim -3.5$ \citep{ji2020car}, none of the UMP stars discovered in Sculptor dSph and the LMC are CEMP \citep{skuladottir2021, chiti2024lmc, limberg2025, ji2025}. On the other hand, the only UMP star known in any UFD \citep[in Pictor~II, an LMC satellite;][]{Drlica-Wagner2016pic2, pace2025pic2} is extraordinarily carbon-enhanced \citep[$\cfe \gtrsim 3.2$;][]{chiti2025}.
 

The currently-accepted understanding of the above-described phenomenon is that, due to their low masses and weak gravitational potential wells, UFD galaxies can only retain yields from low-energy supernova explosions \citep{cooke2014, vanni2023}. These low-energy explosions ($E_{\rm SN} < 2\times10^{51}\,{\rm erg}$) are capable of ejecting lighter elements residing in the outer layers of the massive Population~III supernova progenitor (C, N, O), while the heavier elements from the inner regions (e.g., Fe) fall back into the compact stellar remnant \citep[e.g.,][]{nomoto2013}. Indeed, the recently-discovered UMP CEMP star in the Pictor~II UFD galaxy has an estimated upper limit on the progenitor explosion energy of $E_{\rm SN} < 2\times10^{51}\,{\rm erg}$, but an even lower preferred value of $3\times10^{50}\,{\rm erg}$ assuming a fiducial value of $\rm[Ca/Fe] = +0.4$ \citep{chiti2025}.

Contrary to UFDs, the more massive halos hosting dSph galaxies, and even brighter dwarfs such as the LMC, are capable of retaining the material ejected by high-energy supernovae explosions, such as hypernovae ($E_{\rm SN} \gtrsim 5\times10^{51}\,{\rm erg}$), associated with less carbon production \citep{cooke2014, vanni2023}. All the UMP stars in Sculptor dSph and the LMC are not CEMP and, consistently, have estimated Population~III progenitors with high explosion energies \citep[$4 \lesssim E_{\rm SN}/(10^{51}\,{\rm erg}) \leq 10$;][]{skuladottir2021, chiti2024lmc, ji2025}. In this context, since the Helmi Stream is the debris of a disrupted dwarf galaxy similar to surviving dSph systems, \hesstar represents the first CEMP UMP star discovered in a dSph-like environment.

In the spirit of the above discussion, one hypothesis for the presence of \hesstar in the Helmi Stream is that dSph galaxies should also have accreted smaller UFDs during their hierarchical growth. 
This is corroborated by recent results presented in \citet{zhang2024}, who find no CEMP-no stars in the Helmi stream, only carbon-normal stars.
A galaxy with stellar mass similar to Fornax dSph or Sagittarius dSph, hence the ancient Helmi Stream as well, is expected to have merged with many dozens of smaller halos \citep{griffen2018, deason2023} and, therefore, incorporated their second-generation CEMP stars. Indeed, Local Group dSph galaxies do show strong evidence for past interactions from observed stars with UFD-like chemistry in them, with either undetected or extremely low levels of neutron-capture elements \citep{Fulbright2004draco, roederer2023sxt, ou2025sgr}.
Therefore, we conjecture that \hesstar was born in a UFD environment that was accreted by the Helmi Stream progenitor prior to merging with the Milky Way. Within this proposed scenario, one testable prediction is that tidally unscathed dSph galaxies and the LMC must also host UMP CEMP stars from disrupted UFDs. Also, it should be possible to predict the UMP CEMP fraction in the Milky Way from cosmological merger trees combined with galaxy--halo connection prescriptions and chemical evolution models of carbon production by Population~III stars.


\section{Conclusions}
\label{conclusion}

In this work, we presented the discovery of \hesump, a CNO-enhanced UMP star first identified by the Hamburg/ESO Survey. At $\metal = -4.11$, this is the first UMP star found to be associated with any stellar stream. The orbital and dynamical parameters for \hesstar strongly suggest an association with the Helmi Stream disrupted dwarf galaxy.
Chemical abundances for 18 elements were determined and are consistent with other stars in the same metallicity regime. By comparing the light-element abundance pattern with yields from Population~III faint supernova models, we infer that \hesump\, is a \textit{bona fide} second-generation, mono-enriched, star formed from a gas cloud enriched by a single low-energy 50\,\Msun Population~III star explosion in the early universe. Finally, \hesstar is the first UMP CEMP star identified to be associated with a dSph-like environment. Nevertheless, we conjecture that \hesstar was originally born in a UFD galaxy accreted by the Helmi Stream progenitor system prior to merging with the Milky Way.

\begin{acknowledgments}

The work of V.M.P. is supported by NOIRLab, which is managed by the Association of Universities for Research in Astronomy (AURA) under a cooperative agreement with the U.S. National Science Foundation. G.L. acknowledges support from KICP/UChicago through a KICP Postdoctoral Fellowship. 
This research has made use of NASA's Astrophysics Data System Bibliographic
Services; the arXiv pre-print server operated by Cornell University; the
{\texttt{SIMBAD}} database hosted by the Strasbourg Astronomical Data Center;
and the online Q\&A platform {\texttt{stackoverflow}}
(\href{http://stackoverflow.com/}{http://stackoverflow.com/}).
This work has made use of data from the European Space Agency (ESA) mission Gaia (\url{https://www.cosmos.esa.int/gaia}), processed by the Gaia Data Processing and Analysis Consortium (DPAC, \url{https://www.cosmos.esa.int/web/gaia/dpac/consortium}). Funding for the DPAC has been provided by national institutions, in particular the institutions participating in the Gaia Multilateral Agreement.

\end{acknowledgments}

\software{
{\texttt{Astropy}}\,\citep{astropy2013,astropy2018}, 
{\texttt{awk}}\,\citep{awk}, 
{\texttt{CMasher}}\,\citep{cmasher}, 
{\texttt{dustmaps}}\,\citep{green2018}, 
{\texttt{gnuplot}}\,\citep{gnuplot}, 
{\texttt{jupyter}}\,\citep{jupyter2016}.
{\texttt{NOIRLab IRAF}}\,\citep{tody1986,tody1993,fitzpatrick2025}, 
{\texttt{linemake}}\,\citep{placco2021,placco2021a},
{\texttt{matplotlib}}\,\citep{matplotlib}, 
{\texttt{MOOG}}\,\citep{sneden1973},  
{\texttt{numpy}}\,\citep{numpy}, 
{\texttt{pandas}}\,\citep{pandas}, 
{\texttt{RVSearch}}\,\citep{Rosenthal2021}, 
{\texttt{sed}}\,\citep{sed},
{\texttt{stilts}}\,\citep{stilts},
{\texttt{TOPCAT}}\,\citep{TOPCAT2005}.
}

\facilities{Magellan:Clay (MIKE), NTT (EFOSC2)}

\vfill
\newpage

\bibliographystyle{aasjournal}
\bibliography{bibliografia}

\twocolumngrid

\startlongtable

\appendix
\restartappendixnumbering

\section{Atomic line list}

\begin{deluxetable}{lrrrrrrrrrrrrrrr} 
\tabletypesize{\tiny}
\tabletypesize{\footnotesize}
\tablewidth{0pc}
\tablecaption{\label{eqwl} Atomic Data and Derived Abundances}
\tablehead{
\colhead{Ion}&
\colhead{$\lambda$}&
\colhead{$\chi$} &
\colhead{$\log\,gf$}&
\colhead{$EW$}&
\colhead{$\log\epsilon$\,(X)}&
\colhead{$\Delta$NLTE}\\
\colhead{}&
\colhead{({\AA})}&
\colhead{(eV)} &
\colhead{}&
\colhead{(m{\AA})}&
\colhead{}&
\colhead{}}
\startdata
CH (C)      & 4300.00 & \nodata & \nodata &    syn &    6.520 & \nodata \\
CH (C)      & 4322.00 & \nodata & \nodata &    syn &    6.520 & \nodata \\
NH (N)      & 3360.00 & \nodata & \nodata &    syn &    6.480 & \nodata \\
CN (N)      & 3883.00 & \nodata & \nodata &    syn &    6.580 & \nodata \\
\ion{O}{1}  & 7774.17 &    9.15 &    0.20 &    syn &    7.190 & \nodata \\
\ion{Na}{1} & 5889.95 &    0.00 &    0.11 &  24.65 &    1.711 &   0.038 \\
\ion{Na}{1} & 5895.92 &    0.00 & $-$0.19 &  18.52 &    1.860 &   0.038 \\
\ion{Mg}{1} & 5172.68 &    2.71 & $-$0.36 &  82.60 &    3.631 &   0.265 \\
\ion{Mg}{1} & 5183.60 &    2.72 & $-$0.17 &  92.77 &    3.604 &   0.250 \\
\ion{Al}{1} & 3944.00 &    0.00 & $-$0.64 &    syn &    1.450 & \nodata \\
\ion{Al}{1} & 3961.52 &    0.01 & $-$0.33 &    syn &    1.590 & \nodata \\
\ion{Si}{1} & 3905.52 &    1.91 & $-$1.04 &    syn &    3.260 &   0.090 \\
\ion{Ca}{1} & 4226.74 &    0.00 &    0.24 &    syn &    2.190 &   0.350 \\
\ion{Sc}{2} & 4246.82 &    0.32 &    0.24 &    syn & $-$1.150 & \nodata \\
\ion{Ti}{2} & 4395.03 &    1.08 & $-$0.54 &  42.36 &    1.111 &   0.070 \\
\ion{Ti}{2} & 4417.71 &    1.17 & $-$1.19 &  15.31 &    1.284 &   0.112 \\
\ion{Ti}{2} & 4443.80 &    1.08 & $-$0.71 &  42.15 &    1.272 &   0.009 \\
\ion{Ti}{2} & 4450.48 &    1.08 & $-$1.52 &   9.19 &    1.264 &   0.084 \\
\ion{Ti}{2} & 4501.27 &    1.12 & $-$0.77 &  33.34 &    1.222 &   0.118 \\
\ion{Ti}{2} & 4533.96 &    1.24 & $-$0.53 &  44.01 &    1.293 &   0.131 \\
\ion{Ti}{2} & 4571.97 &    1.57 & $-$0.31 &  37.34 &    1.332 &   0.034 \\
\ion{Ti}{2} & 5188.69 &    1.58 & $-$1.05 &   6.99 &    1.174 &   0.015 \\
\ion{Cr}{1} & 4274.80 &    0.00 & $-$0.22 &    syn &    1.240 &   0.923 \\
\ion{Cr}{1} & 4289.72 &    0.00 & $-$0.37 &    syn &    1.110 &   0.923 \\
\ion{Mn}{1} & 4030.75 &    0.00 & $-$0.50 &    syn &    0.580 &   0.852 \\
\ion{Mn}{1} & 4033.06 &    0.00 & $-$0.65 &    syn &    0.680 &   0.848 \\
\ion{Fe}{1} & 3440.99 &    0.05 & $-$0.96 &  99.23 &    3.348 &   0.389 \\
\ion{Fe}{1} & 3618.77 &    0.99 & $-$0.00 & 101.43 &    3.406 &   0.334 \\
\ion{Fe}{1} & 3758.23 &    0.96 & $-$0.01 & 109.58 &    3.373 &   0.429 \\
\ion{Fe}{1} & 3787.88 &    1.01 & $-$0.84 &  78.07 &    3.479 &   0.422 \\
\ion{Fe}{1} & 3899.71 &    0.09 & $-$1.52 &  85.71 &    3.265 &   0.407 \\
\ion{Fe}{1} & 3922.91 &    0.05 & $-$1.63 &  86.58 &    3.342 &   0.409 \\
\ion{Fe}{1} & 4005.24 &    1.56 & $-$0.58 &  53.89 &    3.359 &   0.388 \\
\ion{Fe}{1} & 4045.81 &    1.49 &    0.28 & 101.94 &    3.369 &   0.423 \\
\ion{Fe}{1} & 4071.74 &    1.61 & $-$0.01 &  85.57 &    3.408 &   0.412 \\
\ion{Fe}{1} & 4132.06 &    1.61 & $-$0.68 &  46.19 &    3.376 &   0.391 \\
\ion{Fe}{1} & 4143.87 &    1.56 & $-$0.51 &  55.47 &    3.296 &   0.399 \\
\ion{Fe}{1} & 4202.03 &    1.49 & $-$0.69 &  52.03 &    3.337 &   0.391 \\
\ion{Fe}{1} & 4250.79 &    1.56 & $-$0.71 &  50.49 &    3.406 &   0.384 \\
\ion{Fe}{1} & 4260.47 &    2.40 &    0.08 &  40.16 &    3.376 &   0.375 \\
\ion{Fe}{1} & 4325.76 &    1.61 &    0.01 &  87.41 &    3.376 &   0.410 \\
\ion{Fe}{1} & 4404.75 &    1.56 & $-$0.15 &  81.22 &    3.345 &   0.451 \\
\ion{Fe}{1} & 4415.12 &    1.61 & $-$0.62 &  56.83 &    3.452 &   0.419 \\
\ion{Fe}{1} & 5192.34 &    3.00 & $-$0.42 &   5.04 &    3.405 &   0.403 \\
\ion{Fe}{1} & 5324.18 &    3.21 & $-$0.11 &   5.13 &    3.325 &   0.391 \\
\ion{Fe}{1} & 5341.02 &    1.61 & $-$1.95 &   5.14 &    3.389 & \nodata \\
\ion{Fe}{1} & 5371.49 &    0.96 & $-$1.64 &  42.61 &    3.440 &   0.441 \\
\ion{Fe}{1} & 5397.13 &    0.92 & $-$1.98 &  28.18 &    3.478 &   0.445 \\
\ion{Fe}{1} & 5405.77 &    0.99 & $-$1.85 &  32.14 &    3.503 &   0.444 \\
\ion{Fe}{1} & 5429.70 &    0.96 & $-$1.88 &  32.69 &    3.508 &   0.441 \\
\ion{Fe}{1} & 5434.52 &    1.01 & $-$2.13 &  15.22 &    3.400 &   0.446 \\
\ion{Fe}{2} & 4923.92 &    2.89 & $-$1.26 &  23.88 &    3.375 &   0.019 \\
\ion{Fe}{2} & 5018.43 &    2.89 & $-$1.10 &  34.53 &    3.424 & \nodata \\
\ion{Co}{1} & 3995.31 &    0.92 & $-$0.18 &  16.56 &    1.361 & \nodata \\
\ion{Ni}{1} & 3783.53 &    0.42 & $-$1.40 &  27.92 &    2.173 & \nodata \\
\ion{Ni}{1} & 3807.14 &    0.42 & $-$1.23 &  50.71 &    2.409 & \nodata \\
\ion{Zn}{1} & 4722.16 &    4.03 & $-$0.37 &    syn &    1.560 & \nodata \\
\ion{Sr}{2} & 4077.71 &    0.00 &    0.15 &    syn & $-$1.580 & \nodata \\
\ion{Sr}{2} & 4215.52 &    0.00 & $-$0.17 &    syn & $-$1.500 & \nodata \\
\ion{Ba}{2} & 4554.03 &    0.00 &    0.16 &    syn & $-$2.820 & \nodata \\

\enddata
\end{deluxetable}

\section{Systematic Abundance Uncertainties}

\begin{deluxetable}{@{}lrrrrr@{}}[!ht]
\tabletypesize{\small}
\tabletypesize{\footnotesize}
\tablewidth{0pc}
\tablecaption{Systematic Abundance Uncertainties \label{sys}}
\tablehead{
\colhead{Ion}&
\colhead{$\Delta$\teff}&
\colhead{$\Delta$\logg}&
\colhead{$\Delta\xi$}&
\colhead{$\sigma$}&
\colhead{$\sigma_{\rm tot}$\vv}\\
\colhead{}&
\colhead{$+$100\,K}&
\colhead{$+$0.20 dex}&
\colhead{$+$0.15 \kmsec}&
\colhead{}&
\colhead{}}
\startdata
\ion{Na}{1} &    0.08 & $-$0.02 & $-$0.01 &    0.10 &    0.13 \\
\ion{Mg}{1} &    0.09 & $-$0.01 & $-$0.04 &    0.10 &    0.14 \\
\ion{Ti}{2} &    0.05 &    0.06 &    0.00 &    0.10 &    0.13 \\
\ion{Fe}{1} &    0.12 & $-$0.02 & $-$0.04 &    0.10 &    0.16 \\
\ion{Fe}{2} &    0.01 &    0.06 & $-$0.01 &    0.10 &    0.12 \\
\ion{Co}{1} &    0.12 & $-$0.01 &    0.00 &    0.15 &    0.19 \\
\ion{Ni}{1} &    0.12 & $-$0.02 & $-$0.01 &    0.10 &    0.16 \\
\enddata
\tablenotetext{a}{Calculated from the quadratic sum of the individual error estimates.}
\end{deluxetable}

\end{document}